\documentclass[reprint, superscriptaddress, amsmath, amssymb, aps]{revtex4-1}
\usepackage{epstopdf}
\usepackage{graphicx}
\usepackage{dcolumn}
\usepackage{bm}
\usepackage{multirow}
\usepackage[table,xcdraw]{xcolor}
\usepackage{color,soul}

\begin{document}
\title{Carrier mobilities of Janus transition metal dichalcogenides monolayers studied by Born effective charge and first-principles calculation}
\author{Jingxin Hu}
\author{Kui Rao}
\author{Jing Luo}
\author{Lianjie Hu}
\affiliation{Department of Physics and Key Laboratory for Low-Dimensional Structures and Quantum Manipulation (Ministry of Education), Hunan Normal University, Changsha 410081, China}
\author{Ziran Liu}
\email{zrliu@hunnu.edu.cn}
\affiliation{Department of Physics and Key Laboratory for Low-Dimensional Structures and Quantum Manipulation (Ministry of Education), Hunan Normal University, Changsha 410081, China}
\affiliation{Key Laboratory for Matter Microstructure and Function of Hunan Province, Hunan Normal University, Changsha 410081, China}

\date{\today}
\begin{abstract}
Two-dimensional (2D) Janus transition metal dichalcogenides (TMDs) are a new class of materials with unique physical properties. However, the carrier mobility of most Janus TMDs calculated by deformation potential theory (DPT) is not reliable due to the unconsidered part of lattice scattering. In this work, we propose a new method of Born effective charge (BEC) to calculate the carrier mobility of Janus TMDs by including the important factors that neglected in the DPT. The BEC could be used in the calculation of both pure and defective Janus TMDs by employing density functional perturbation theory. We have figured out the relationship between the carrier mobility and the value of BEC, which is the lower the absolute BEC, the higher the electron or hole mobility. Using the new method, we have calculated the carrier mobility of commonly studied Janus TMDs with and without defect. The method may shed light on the high-throughout calculation of selecting high carrier mobility 2D materials.
\end{abstract}

\maketitle

\section{INTRODUCTION}
Two-dimensional TMDs crystals, such as H-MoS$_{2}$ and H-MoTe$_{2}$ have been extensively studied by experiment and first-principles calculations \cite{jin2018prediction,shi2018mechanical,xia2018universality,liu2019strain,guo2018biaxial,dong2017large}. Two-dimensional (2D) Janus materials are a new class of materials with unique physical, chemical, and quantum properties \cite{yagmurcukardes2020quantum}. Janus TMDs of MoSSe was firstly synthesized and confirmed by means of scanning transmission electron microscopy and energy-dependent X-ray photoelectron spectroscopy \cite{lu2017janus,zhang2017janus}. The structure of MoSSe was based on single-crystalline triangular MoS$_{2}$ monolayers, through chemical vapour deposition, Se atoms replace the S atoms, forming a structurally stable Janus MoSSe monolayer in which the Mo atoms are covalently bonded to underlying S and top-layer Se atoms.

From then on, the interesting Janus 2D materials attracting a lot of attentions \cite{sant2020synthesis, ghobadi2020electrical, paez2021stability, xu2021colossal, petric2021raman}. Janus WSSe has been synthesized and it's novel physical properties were also revealed \cite{trivedi2020room,lin2020low}. Then the VSSe \cite{zhang2019first}, the ZrSSe \cite{vu2019electronic}, and the SnSSe \cite{guo2019predicted} have been systematically studied, and a lot of interesting intrinsic physical properties were revealed.
Single-layer PtSSe exhibits a high absorption coefficient in the visible light region, appropriate band edge positions and strong ability for carrier separation and transfer \cite{peng2019two}. Besides the experiments, first-principles calculations have also been applied to study the stabilities of the Janus structures \cite{kandemir2018janus,cheng2013spin}. Obviously, the symmetry of Janus structure is lower than 2D TMDs materials, which indicate that the multi-layer Janus TMDs may have high out-of-plane piezoelectric effect \cite{dong2017large,yagmurcukardes2019electronic}. The electron and hole mobilities of Janus TMDs are important descriptors for selecting good candidates for advanced 2D materials design.

The deformation potential theory (DPT) is frequently used to calculate the intrinsic mobility of semiconductors \cite{bardeen1950deformation,xi2012first,qiao2014high,yang2016two,cai2014polarity,zhu2017multivalency,long2009theoretical} and the Janus 2D materials \cite{yin2018tunable,patel2020high,guo2019predicted}. However, the application of DPT must satisfy the following requirements \cite{xi2012first,price1981two}. One is that the lowest valence band of the material should be parabolic, and the other one is about the scattering of the carrier, which should be mainly caused by longitudinal acoustic wave of the lattice. Besides, to obtain the carrier mobility in DPT, the effective mass should be calculated from averaging all directions of materials. Therefore, using DPT to calculate carrier mobility of anisotropic materials is challengeable.

As we all know, electron-phonon coupling (EPC) is the relative accurate method to calculate the carrier mobility of 2D materials \cite{kaasbjerg2012phonon,li2015electrical,gunst2016first,hinsche2017spin}. However, the calculation of EPC is very expensive, and the cell is limited within ten atoms without defects. Therefore, it is necessary to find a new way to calculate the mobility effectively, especially for the structure with defects. The BEC can affect the vibration of the longitudinal optical branch which cause a considerable change in polarization, resulting in stronger carrier scattering through coulomb interaction, thus has been applied to predict the carrier mobility of a amount of 2D TMDs \cite{cheng2018limits}. BEC can be calculated by applying changes in polarization divided by the displacement of the atoms. In this work, we use first-principles density functional perturbation theory to calculate BEC, then predict the carrier mobility of perfect and defective hexagonal Janus structure (H-) of H-MoSSe, H-MoSTe, H-MoSeTe, H-WSSe, H-WSTe, H-WSeTe and triangular Janus structure (T-) of T-HfSSe, T-ZrSSe, T-SnSSe, T-PtSSe, T-PtSTe, T-PtSeTe.

\section{Methods}

\begin{figure}
  \centering
  \includegraphics[width=0.45 \textwidth]{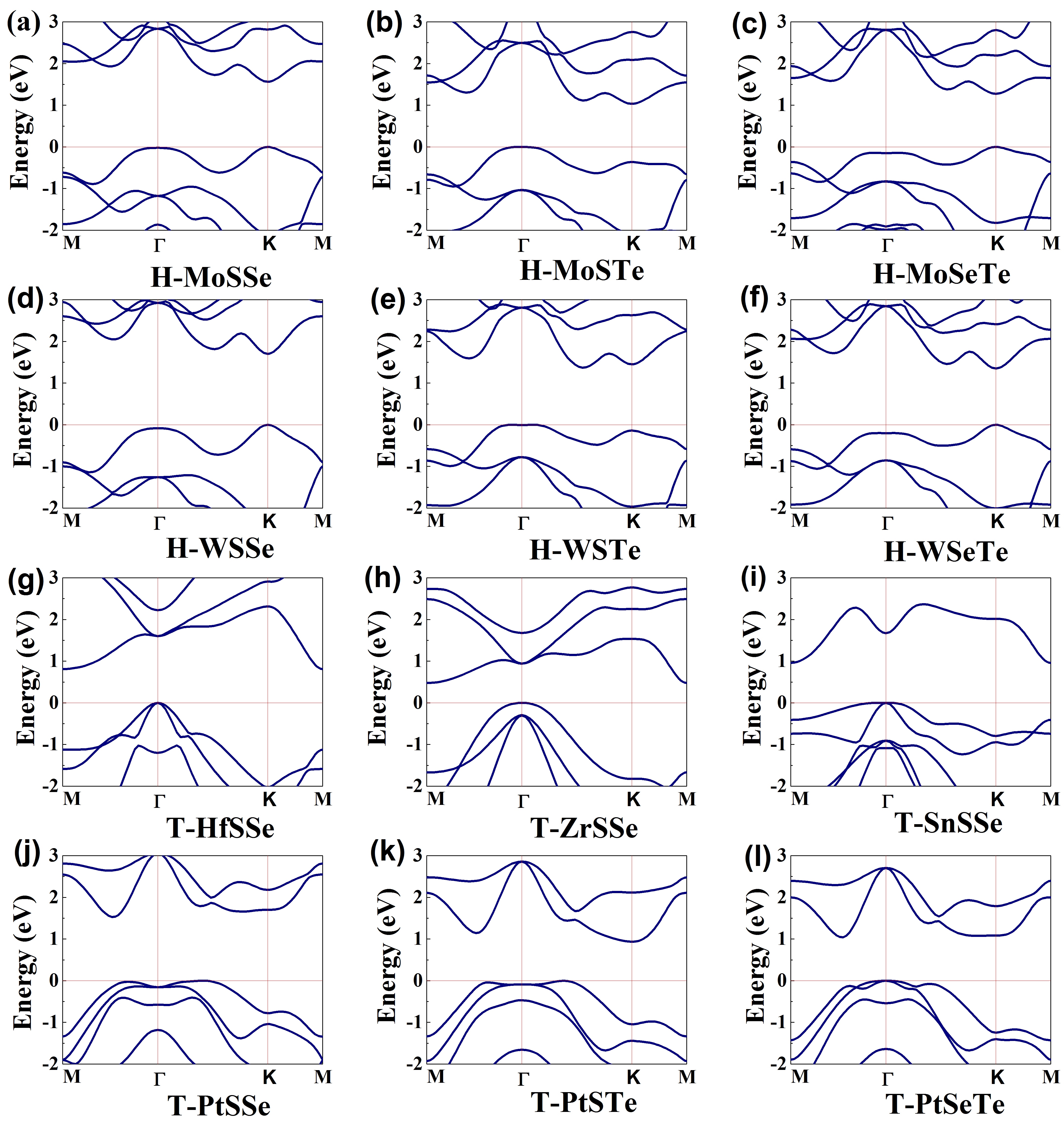}\\
  \caption{The band structures of H (a-f) and T(g-l) types of Janus TMDs monolayers. The Fermi level is set to zero.}\label{bandstructure-janus}
\end{figure}

The calculations of band structure and carrier mobility are based on density functional theory (DFT) using the plane-wave basis within projection augmented wave (PAW) method. The generalized gradient approximation (GGA) in the Perdew-Burke Ernzerhof (PBE) \cite{perdew1996generalized,perdew1997generalized} form is used as the exchange correlation potential implemented in the Vienna ab initio simulation package (VASP) \cite{kresse1993ab,kresse1994ab}. On the premise of ensuring the accuracy of the calculation results, we set the cutoff energy of all related calculations to 350 eV, the convergence criterion between the two electronic steps in the self-consistent process to $10^{-6}$ eV and the force convergence is set to 0.01 eV/{\AA}, the K-points are set to 9$\times$9$\times$1 and 15$\times$11$\times$1 for relaxation and energy calculation, respectively.

We obtain the phonon spectrum using the density functional perturbation theory (DFPT) method \cite{baroni2001phonons} as implemented in VASP and post-processed by using the Phonopy package \cite{togo2015first}. In order to improve the precise of the calculation, we increase the energy convergence criteria for electronic and force convergence, which were set to be $10^{-10}$ eV and 0.001 eV/{\AA}, respectively. A vacuum layer with a thickness of 20{\AA} is added along the z-axis in order to avoid interaction between adjacent layers.

The phonon-limited formula is used to estimate carrier mobility in low dimensional materials \cite{xi2012first,bruzzone2011ab,takagi1994universality}:
\begin{equation}
 \mu = \frac {e\hbar^{3} C_{2D}}{K_{B}Tm^{\ast}m_d E_i^{2}} m_{d}= \sqrt{m_x^{\ast}m_y^{\ast}}
\label{equ1}
\end{equation}
where $m^*$ is the effective mass along the transmission direction ($m^*_x$ or $m^*_y$ is along $x$ or $y$ direction, respectively), and the mean effective mass was defined as $m_d=\sqrt{m_{x}^{*}m_{y}^{*}}$. $C_{2D}$ is the elastic modulus of the simplex deformed crystal, which is used to simulate the lattice distortion caused by strain, which is given by $C_{2D}=\dfrac{1}{S_0} \dfrac{\partial^2 E}{\partial(\Delta l/ l_0)^2} \vert_{l=l_0}$, $E$ is the total energy of the system and $S_0$ is the equilibrium area of the unit cell, $l_0$ is the lattice constant, $\Delta l= l-l_0$ is corresponding lattice distortion. Here the applied deformation varies from 0 to 0.6$\%$. $E_i$ is the deformation potential, which can be expressed as the following formula $E= \dfrac{\partial E_{edge}}{\partial(\Delta l/l_0)}$, which represents of the strain-induced band edge displacement (the minimum electron-conducting band, the maximum hole valence band) along the direction of electron transport. The temperature used for the mobility calculations was 300 K.

Density functional perturbation theory (DFPT) is used to calculate the BEC of two-dimensional Janus structures. The BEC of an atom $s$ is defined as the change in polarization $p$ in the $\alpha$ direction linearly induced by a sublattice displacement $u_s$ in the $\beta$ direction in zero macroscopic electric field as shown in below formula \cite{1992Theory}:
\begin{equation}
Z_s^{*\alpha,\beta}= V \frac{\partial P_{\alpha}}{\partial u_{s,\beta}},
\end{equation}
where $V$ is the unit volume, $P_\alpha$ is the $\alpha$ component of the polarization and $u_{s,\beta}$ is the displacement of the $s^{th}$ atom in the direction of $\beta$. For two dimensional Janus structures with low symmetry, BEC is not symmetric in the Cartesian directions. Since any rigid translation of the entire system does not cause macroscopic polarization, the BEC tensor follows acoustics and rules. The sum rule thus was given as \cite{1992Theory,1970Microscopic}:
\begin{equation}
\sum_{s}Z_s^{*\alpha,\beta}= 0
\end{equation}
i.e., the sum of BEC of all the atoms in certain system is zero.

\section{RESULTS AND DISCUSSION}
\subsection{The failure of the mobility from DPT}

\begin{table*}[ht]
\caption{The electron mobility $(\mu_e)$ of T- and H-TMDs calculated by employing DPT.}
\renewcommand{\arraystretch}{1.2}
\setlength{\tabcolsep}{2mm}{ 
\small
\begin{tabular}{|l|c|c|}
\hline
Systems            & Present work ($\mu_e$)(cm$^{2}$/Vs)                       & Experimental results available ($\mu_e$)(cm$^{2}$/Vs)            \\\hline
T-ZrS$_{2}$       &  $\mu$$_{x}$=261.27 ,  $\mu$$_{y}$=549.32                       & \                                       \\ \hline
T-ZrSe$_{2}$      &  $\mu$$_{x}$=1252.70,  $\mu$$_{y}$=2549.70                         & \                                      \\ \hline                                                                                                                      T-HfS$_{2}$       &  $\mu$$_{x}$=428.77 ,  $\mu$$_{y}$=1110.00                       & \                                   \\ \hline
T-HfSe$_{2}$      &  $\mu$$_{x}$=2763.00,  $\mu$$_{y}$=7323.00                           & 0.22 \cite{kang2015electrical}              \\ \hline
T-SnS$_{2}$       &  $\mu$$_{x}$=156.58 ,  $\mu$$_{y}$=491.98                                   &  \                                  \\ \hline                                                                                T-SnSe$_{2}$      &  $\mu$$_{x}$=423.60 ,  $\mu$$_{y}$=312.33                                      & 8.6 \cite{su2013snse2}                    \\ \hline                                                                           T-PtS$_{2}$       &  $\mu$$_{x}$=392.12 ,  $\mu$$_{y}$=254.06                                       &  \                                   \\ \hline                                                                                                              T-PtSe$_{2}$      &  $\mu$$_{x}$=884.47 ,  $\mu$$_{y}$=459.92                                        & 18 \cite{yang2021high}                      \\ \hline                                                                                                                     T-PtTe$_{2}$      &  $\mu$$_{x}$=1021.00,  $\mu$$_{y}$=1204.00                                         &  \                                   \\ \hline
H-MoS$_{2}$       &  $\mu$$_{x}$=169.47 ,  $\mu$$_{y}$=165.44                                          & 200 \cite{radisavljevic2011single}      \\\hline                                                                                                                                          H-MoSe$_{2}$      &  $\mu$$_{x}$=110.16 ,  $\mu$$_{y}$=108.70                                       & 50 \cite{wang2014chemical}              \\ \hline
H-MoTe$_{2}$      &  $\mu$$_{x}$=101.55 ,  $\mu$$_{y}$=98.32                                           & 40 \cite{keum2015bandgap}                 \\ \hline
H-WS$_{2}$        &  $\mu$$_{x}$=277.21 ,  $\mu$$_{x}$=272.05                                            & 100 \cite{wang2021electron}               \\ \hline
H-WSe$_{2}$       &  $\mu$$_{x}$=278.00 ,  $\mu$$_{x}$=266.33                                         & 142 \cite{liu2013role}                     \\\hline
\end{tabular}}\label{table-DPTmobility}
\end{table*}

From previous work, we could find that the CBM of TMDs satisfy the parabolic requirement \cite{cheng2018limits}. However, most existing electron mobilities of TMDs predicted by DPT are much larger than the available experimental results, except for MoS$_{2}$ as shown in Table \ref{table-DPTmobility}. From the calculations, we found that the anisotropy of the electron mobilities of TMDs can't be neglected, especially for that of the T-phase of TMDs. However, the lower mobility of two directions (x and y) still much higher than the related experimental results. Hence, some important factors should be considered in to the method of DPT, which at present is not good enough for the electron mobility calculations of the majority of TMDs.

\begin{figure}
  \centering
  \includegraphics[width=0.45 \textwidth]{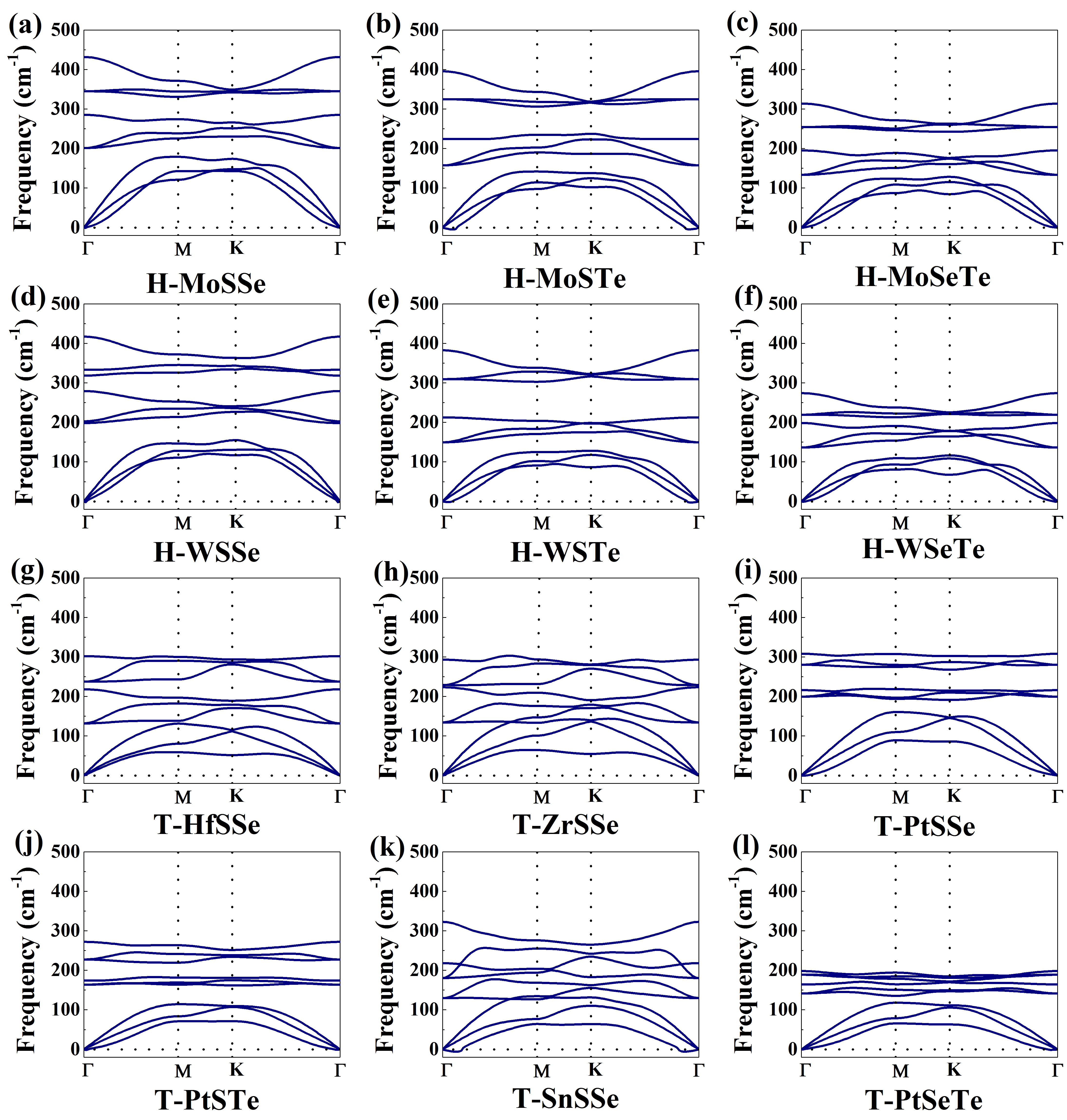}\\
  \caption{The phonon spectrum of H (a-f) and T(g-l) types of Janus TMDs monolayers.}\label{Phonon-janus}
\end{figure}

\begin{figure}
  \centering
  \includegraphics[width=0.45 \textwidth]{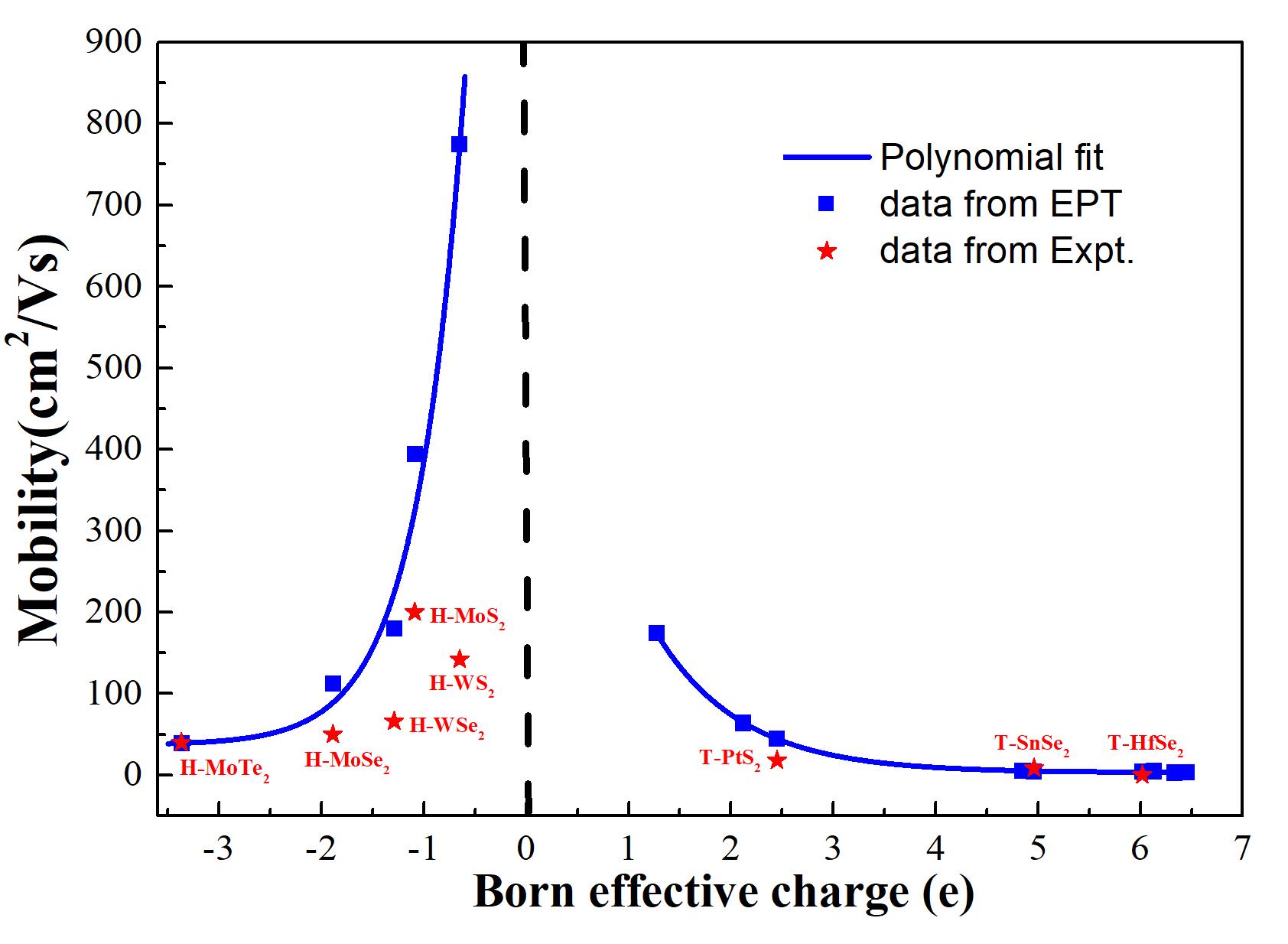}\\
  \caption{The relationship between the BEC and electron mobility. The blue squares represent values calculated by EPC which are fitted by blue lines. The red five-pointed stars are available experimental values.}\label{born-mobility-ele}
\end{figure}

How about the electron mobilities of Janus TMDs from DPT? Table \ref{table-janus-dpt} shows that the electron mobilities of Janus TMDs from DPT are still not satisfactory for describing the affects from lattice scattering, i.e., the electron mobilities are unreasonably high although the band structure of the system satisfies the parabolic properties, as shown in Fig. \ref{bandstructure-janus}. In order to get reliable data of carrier mobility of Janus TMDs, new methods or factors should be employed in the calculations.

\begin{table*}[ht]
\caption{The calculated electron $(\mu_e)$ and hole $(\mu_h)$ mobility of Janus TMDs from DPT.}
\renewcommand{\arraystretch}{1.4}
\setlength{\tabcolsep}{2mm}{ 
\small
\begin{tabular}{|l|c|c|c|}
\hline
System            & $(\mu_e)$(cm$^{2}$/Vs)                              & $(\mu_h)$(cm$^{2}$/Vs)                              \\\hline
T-ZrSSe           &  $\mu$$_{x}$=1971.80 , $\mu$$_{y}$=3646.70          &  $\mu$$_{x}$=495.48,  $\mu$$_{y}$=500.73                            \\ \hline
T-HfSSe           &  $\mu$$_{x}$=2033.40 , $\mu$$_{y}$=4864.50          &  $\mu$$_{x}$=452.84,  $\mu$$_{y}$=487.95                       \\ \hline
T-SnSSe           &  $\mu$$_{x}$=231.91  , $\mu$$_{y}$=883.48           &  $\mu$$_{x}$=0.74  ,  $\mu$$_{y}$=0.19                          \\ \hline
T-PtSSe           &  $\mu$$_{x}$=707.26  , $\mu$$_{y}$=377.51           &  $\mu$$_{x}$=17.71 ,  $\mu$$_{y}$=407.89                               \\ \hline
T-PtSTe           &  $\mu$$_{x}$=59.28   , $\mu$$_{y}$=51.57            &  $\mu$$_{x}$=0.64  ,  $\mu$$_{y}$=1.40                          \\ \hline
T-PtSeTe          &  $\mu$$_{x}$=389.28  , $\mu$$_{y}$=590.72           &  $\mu$$_{x}$=14.23 ,  $\mu$$_{y}$=14.33                           \\ \hline
H-MoSSe           &  $\mu$$_{x}$=120.38  , $\mu$$_{y}$=117.25           &  $\mu$$_{x}$=233.23,  $\mu$$_{y}$=236.76                      \\\hline
H-MoSTe           &  $\mu$$_{x}$=46.53   , $\mu$$_{y}$=45.41            &  $\mu$$_{x}$=2.25  ,  $\mu$$_{y}$=1.91                           \\ \hline
H-MoSeTe          &  $\mu$$_{x}$=68.90   , $\mu$$_{y}$=68.91            &  $\mu$$_{x}$=87.62 ,  $\mu$$_{y}$=90.92                               \\ \hline
H-WSSe            &  $\mu$$_{x}$= 267.31 , $\mu$$_{y}$=261.47           &  $\mu$$_{x}$=513.85,  $\mu$$_{y}$=533.27                         \\ \hline
H-WSTe            &  $\mu$$_{x}$= 158.94 , $\mu$$_{y}$=145.67           &  $\mu$$_{x}$=15.14 ,  $\mu$$_{y}$=15.29                           \\\hline
H-WSeTe           &  $\mu$$_{x}$=183.50  , $\mu$$_{y}$=177.66           &  $\mu$$_{x}$=267.66,  $\mu$$_{y}$=273.97                                \\ \hline
\end{tabular}}\label{table-janus-dpt}
\end{table*}

\begin{figure}
  \centering
  \includegraphics[width=0.45 \textwidth]{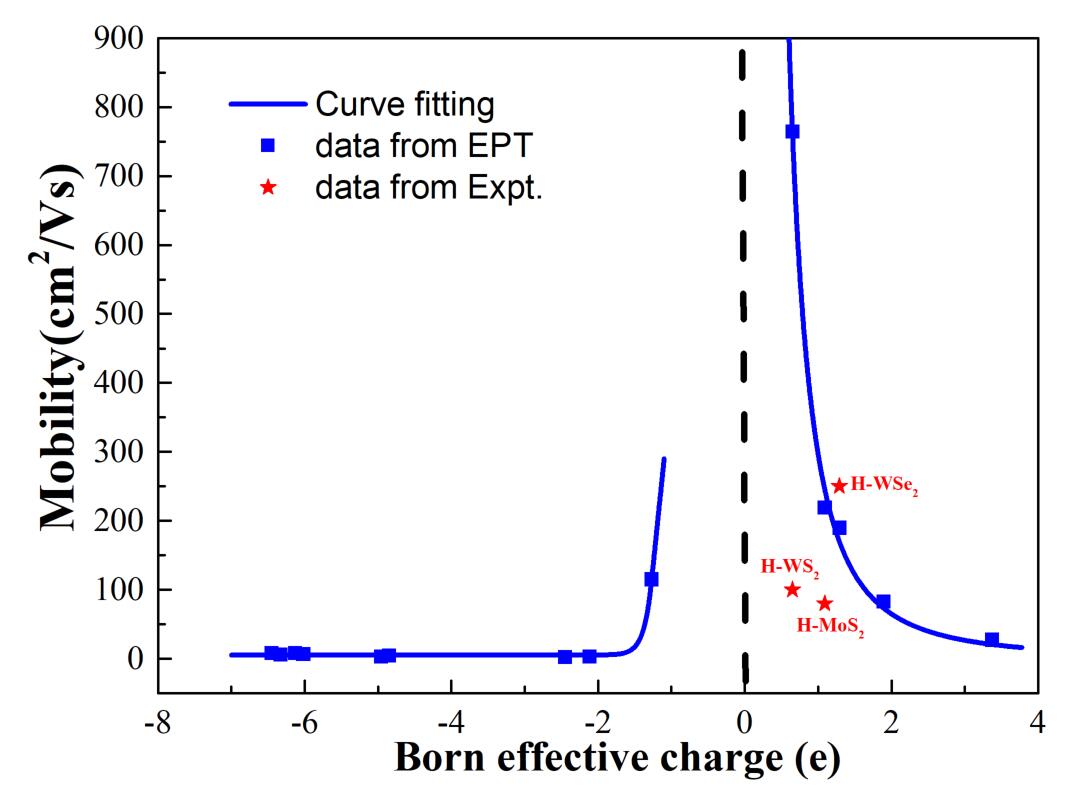}\\
  \caption{The relationship between the BEC and hole mobility. The blue squares represent values calculated by EPC which are fitted by blue lines. The red five-pointed stars are available experimental values.}\label{born-mobility-hole}
\end{figure}

\subsection{The bridge between mobility and BEC}
As we all know, EPC is a relatively precise way to calculate the carrier mobility recently. However, this method is very expensive and not suitable for a large system not to mention the system with defects. Hence, finding a new way to describe the main contribution from lattice scattering is necessary for building a method for high-throughout calculation. BEC is such a quantity that could describe the lattice scattering effectively and then obtain the carrier mobility.

The electron mobility of TMDs from both experiment and EPC are summarized in Fig. \ref{born-mobility-ele}. After fitting the data from EPC, we find that the electron mobility from EPC has polynomial function with BEC in system of TMDs. Besides, the experimental results approximately meet the fitting values. It is easy to conclude that the lower the absolute BEC, the higher the electron mobility, which is similar to that of hole mobility, as shown in Fig. \ref{born-mobility-hole}. In this bridge, the carrier mobility of the TMDs can be obtained through BEC.

Then let's promote the method to Janus TMDs which require the bandstructure and phonon spectrum of Janus TMDs have similar characters as that of TMDs. From Fig. \ref{bandstructure-janus} and Fig. \ref{Phonon-janus}, we could know that the bandstructure and phonon spectrum of Janus TMDs do like that of TMDs \cite{cheng2018limits}.
Thus, the mobility of Janus structures could be predicted from the BEC safely. The quantitative electron and hole mobility of Janus structures can be found in Table \ref{table-H-T electron mobility} and Table \ref{table-H-T hole mobility}, respectively.

\subsection{Prediction for Janus structure}
In this section, we want to predict the carrier mobility by applying the above functions with the BEC. For H types of Janus TMDs, the order of the values of metal atom for $|Z_B|$ are H-WSTe$<$H-MoSTe$<$H-WSeTe$<$H-MoSeTe, in which H-WSTe has the largest electron mobility of 129.72 cm$^{2}$V$^{-1}$s$^{-1}$, while H-MoSeTe has the lowest electron mobility of 64.79 cm$^{2}$V$^{-1}$s$^{-1}$, as shown in Table \ref{table-H-T electron mobility}. Like the relationship between $|Z_B|$ and the electron mobility for H-phase, the smaller the value of $|Z_B|$ of nonmetal atom, the larger the hole mobility. As shown in Table \ref{table-H-T hole mobility}, H-MoSeTe has the lowest hole mobility of 36.52 cm$^{2}$V$^{-1}$s$^{-1}$ related to the largest $|Z_B|$ of 2.60 $|e|$, while H-WSSe has the largest hole mobility of 382.05 cm$^{2}$V$^{-1}$s$^{-1}$ related to the smallest $|Z_B|$ of 0.89 $|e|$. We further found that for the TMDs of H-MXY (M refers to metal atom, X and Y are distinct element), the H-WXY has larger carrier mobility relative to H-MoXY because W has the highest Pauling electronegativity among all metal elements \cite{cheng2018limits}.

For T-phase of Janus, the order of the values for $|Z_B|$ of metal atom is T-PtSTe$<$T-PtSeTe$<$T-PtSSe$<$T-SnSSe$<$T-HfSSe$<$T-ZrSSe, in which T-PtSTe has the largest electron mobility of 109.38 cm$^{2}$V$^{-1}$s$^{-1}$ relate to the lowest $|Z_B|$ of 1.83 $|e|$, and T-ZrSSe has the lowest electron mobility of 25.98 cm$^{2}$V$^{-1}$s$^{-1}$, as shown in Table \ref{table-H-T electron mobility}. Similarly to electron mobility, the largest hole mobility for T-PtSTe is 6.28 cm$^{2}$V$^{-1}$s$^{-1}$, as shown in Table \ref{table-H-T hole mobility}.

This correlation is understandable, because a large BEC suggests a large change in the polarization, resulting in a stronger carrier scattering through Coulomb interaction. It's worth mentioning that all the calculated BEC for H-phase is different from that of the chemical charge, which the metals has negative $|Z_B|$ but not positive $|Z_B|$. This can be explained from the local changes of polarization around atoms by establish $\pi$ backbonding model \cite{pike2017origin}. By the way, this method is not suitable for calculating the carrier mobility of H-WSSe and H-MoSSe which is mainly influenced by optical branch rather than acoustic branch in phonon spectrum \cite{cheng2018limits}.

\begin{table}[ht]
\caption{The calculated BEC of metal atom ($\emph{Z}$$_{B}$) and the related electron mobility ($\mu_e$) of H and T types of Janus TMDs monolayers.}
\renewcommand{\arraystretch}{1.4}
\setlength{\tabcolsep}{2mm}{ 
\small
\begin{tabular}{|l|c|c|c|c|c|}
\hline
  Janus TMDs       & $\emph{Z}$$_{B}$ (e)       & Mobility $\mu_e$(cm$^{2}$/Vs)                               \\ \hline
   H-MoSSe     & -1.48              & 156.14                                                      \\ \hline
   H-MoSTe     & -2.27              & 78.44                                                     \\ \hline
   H-MoSeTe    & -2.60              & 64.79                                                           \\ \hline
   H-WSSe      & -0.89              & 394.69                                                      \\ \hline
   H-WSTe      & -1.65              & 129.72                                                          \\ \hline
   H-WSeTe     & -2.57              & 65.82                                                             \\ \hline
   T-ZrSSe     & 7.71               & 25.98                                                           \\ \hline
   T-SnSSe     & 4.93               & 33.18                                                              \\ \hline
   T-HfSSe     & 7.24               & 26.65                                                            \\ \hline
   T-PtSSe     & 2.53               & 66.80                                                              \\ \hline
   T-PtSTe     & 1.83               & 109.38                                                             \\ \hline
   T-PtSeTe    & 2.07               & 90.08                                                             \\ \hline

\end{tabular}}\label{table-H-T electron mobility}
\end{table}

\begin{table}[ht]
\caption{The calculated BEC of nonmetal atom ($\emph{Z}$$_{B}$) and the related hole mobility ($\mu_e$) of H and T types of Janus TMDs monolayers.}
\renewcommand{\arraystretch}{1.4}
\setlength{\tabcolsep}{2mm}{ 
\small
\begin{tabular}{|l|c|c|c|c|c|}
\hline
  System           & $\emph{Z}$$_{B}$(e)        & Mobility $\mu_h$(cm$^{2}$/Vs)         \\ \hline
   H-MoSSe         & 1.48              & 125.43                               \\ \hline
   H-MoSTe         & 2.27              & 49.16                               \\ \hline
   H-MoSeTe        & 2.60              & 36.52                                 \\ \hline
   H-WSSe          & 0.89              & 382.05                                \\ \hline
   H-WSTe          & 1.65              & 98.86                                 \\ \hline
   H-WSeTe         & 2.57              & 37.46                                    \\ \hline
   T-ZrSSe         & -7.71             & 5.26                                  \\ \hline
   T-SnSSe         & -4.93             & 5.26                                    \\ \hline
   T-HfSSe         & -7.24             & 5.26                                    \\ \hline
   T-PtSSe         & -2.53             & 5.26                                    \\ \hline
   T-PtSTe         & -1.83             & 6.28                                    \\ \hline
   T-PtSeTe        & -2.07             & 5.35                                     \\ \hline

\end{tabular}}\label{table-H-T hole mobility}
\end{table}

\begin{table*}[ht]
\caption{The calculated TMDs of the electron mobility ($\mu_e$) and BEC of metal atom (Z$_{B}$).}
\renewcommand{\arraystretch}{1.4}
\setlength{\tabcolsep}{2mm}{ 
\small
\begin{tabular}{|l|c|c|c|c|c|}
\hline
System       & Z$_{B}$(e)        & Mobility $\mu_{e}$(cm$^{2}$/Vs)    & Z$_{B}$ (e) from work \cite{cheng2018limits}            & Mobility $\mu_{e}$(cm$^{2}$/Vs) from work \cite{cheng2018limits}       \\ \hline
H-MoS$_{2}$  & 1.00              & 296.00                     &  1.09    & 219.00                                    \\ \hline
H-MoSe$_{2}$ & 1.80              & 81.70      & 1.89     &  83.10                               \\ \hline
H-MoTe$_{2}$     & 3.38              & 20.56        & 3.37     & 27.30                             \\ \hline
H-WS$_{2}$   & 0.52              & 1239.49        & 0.65     & 764.50                                 \\ \hline
H-WSe$_{2}$  & 1.23              & 188.10      & 1.29     &189.90                                  \\ \hline
T-ZrS$_{2}$ & -7.15             & 5.26   & -6.13    & 8.50                                    \\ \hline
T-ZrSe$_{2}$ & -8.17             & 5.26    & -6.45    & 8.20                                    \\ \hline
T-SnS$_{2}$  & -4.90             & 5.26     & -4.85    & 4.60                                 \\ \hline
T-SnSe$_{2}$ & -4.98             & 5.26     & -4.96    & 3.00                                   \\ \hline
T-HfS$_{2}$ & -6.84             & 5.26   & -6.03    & 6.70                                        \\ \hline
T-HfSe$_{2}$ & -7.60             & 5.26    & -6.33    & 6.00                                       \\ \hline
T-PtS$_{2}$ & -2.69             & 5.26   & -2.45    & 1.90                                      \\ \hline
T-PtSe$_{2}$ & -2.35             & 5.27     & -2.12    & 3.20                                                \\ \hline
T-PtTe$_{2}$ & -1.55             & 22.16     & -1.27    & 115.20                                            \\ \hline
\end{tabular}}\label{TMD electron mobility}
\end{table*}

We have calculated the mobilities of 12 types of MXY, unfortunately, we do not find a MXY with a mobility higher than H-WS$_2$. However, we could deduce that the high-mobility 2D semiconductor maybe more likely exist in materials having zero BEC, which could be easily calculated from first-principles calculations. Besides, it is possible to find higher-mobility candidates by including more 2D materials from database into consideration \cite{mounet2018two}.

\begin{table*}[ht]
\caption{The calculated BEC of metal atom (Z$_{B}$) and the related electron mobility ($\mu_e$) of H and T types of Janus TMDs monolayers. The size of supercells used in calculation of the impurity is 3$\times$3.}
\renewcommand{\arraystretch}{1.4}
\setlength{\tabcolsep}{2mm}{ 
\small
\begin{tabular}{|l|c|c|l|c|c|c|}
\hline
Janus TMDs                & Z$_{B}$(e)   & Mobility $\mu_e$(cm$^{2}$/Vs)& Janus TMDs              & Z$_{B}$(e) & Mobility $\mu_e$(cm$^{2}$/Vs)             \\ \hline
H-Mo$_{9}$Se$_{18}$   & -1.84        & 108.43                     & T-Hf$_{9}$Se$_{18}$   & 7.60       & 26.12                                    \\ \hline
H-Mo$_{9}$Se$_{17}$S  & -1.80        & 112.36                     & T-Hf$_{9}$Se$_{17}$S  & 7.58       & 26.15                                    \\ \hline
H-Mo$_{9}$Se$_{17}$Te & -1.92        & 101.30                     & T-Hf$_{9}$Se$_{17}$Te & 7.73       & 25.95                                     \\ \hline
H-Mo$_{9}$Te$_{18}$   & -3.28        & 48.51                      & T-Zr$_{9}$S$_{18}$    & 7.26       & 26.62                                      \\ \hline
H-Mo$_{9}$Te$_{17}$S  & -3.17        & 50.46                      & T-Zr$_{9}$S$_{17}$Se  & 7.30       & 26.55                                       \\ \hline
H-Mo$_{9}$Te$_{17}$Se & -3.20        & 49.90                      & T-Zr$_{9}$S$_{17}$Te  & 7.40       & 26.41                                       \\ \hline
H-W$_{9}$Se$_{18}$    & -1.23        & 216.65                     & T-Zr$_{9}$Se$_{18}$   & 8.17       & 25.43                                       \\ \hline
H-W$_{9}$Se$_{17}$S   & -1.19        & 230.02                     & T-Zr$_{9}$Se$_{17}$S  & 8.13       & 25.48                                       \\ \hline
H-W$_{9}$Se$_{17}$Te  & -1.31        & 193.48                     & T-Zr$_{9}$Se$_{17}$Te & 8.28       & 25.32                                         \\ \hline
T-Sn$_{9}$S$_{18}$    & 4.89         & 33.38                      & T-Pt$_{9}$S$_{18}$    & 2.68       & 62.21                                         \\ \hline
T-Sn$_{9}$S$_{17}$Se  & 4.90         & 33.32                      & T-Pt$_{9}$S$_{17}$Se  & 2.66       & 62.83                                        \\ \hline
T-Sn$_{9}$S$_{17}$Te  & 4.80         & 33.84                      & T-Pt$_{9}$S$_{17}$Te  & 2.66       & 62.83                                        \\ \hline
T-Sn$_{9}$Se$_{18}$   & 4.98         & 32.94                      & T-Pt$_{9}$Se$_{18}$   & 2.36       & 74.15                                         \\ \hline
T-Sn$_{9}$Se$_{17}$S  & 4.98         & 32.94                      & T-Pt$_{9}$Se$_{17}$S  & 2.38       & 73.26                                        \\ \hline
T-Sn$_{9}$Se$_{17}$Te & 4.99         & 32.94                      & T-Pt$_{9}$Se$_{17}$Te & 1.96       & 98.05                                       \\ \hline
T-Hf$_{9}$S$_{18}$    & 6.88         & 27.25                      & T-Pt$_{9}$Te$_{18}$   & 1.56       & 142.63                                      \\ \hline
T-Hf$_{9}$S$_{17}$Se  & 6.91         & 27.20                      & T-Pt$_{9}$Te$_{17}$S  & 1.64       & 131.05                                     \\ \hline
T-Hf$_{9}$S$_{17}$Te  & 7.00         & 27.04                      & T-Pt$_{9}$Te$_{17}$Se & 1.61       & 135.19                                     \\ \hline

\end{tabular}}\label{table-defect electron mobility}
\end{table*}

\subsection{The mobility of TMDs with point defects}
The prediction of mobility of TMDs with point defects by EPC are challengeable due to a large supercell should be employed in the DFT calculations. Herein we try to use the new method to determine the mobility of TMDs with one substitutional point defect due to the formation of Janus TMDs monolayer origins from substituting one chalcogen atom of TMDs.

\begin{table*}[ht]
\caption{Calculated impurities in TMDs(3$\times$3) of the hole mobility ($\mu_h$), BEC of nonmetal atom(Z$_{B}$).}
\renewcommand{\arraystretch}{1.4}
\setlength{\tabcolsep}{2mm}{ 
\small
\begin{tabular}{|l|c|c|l|c|c|}
\hline
System                & Z$_{B}$(e)   & Mobility $\mu_h$(cm$^{2}$/Vs)& System                & Z$_{B}$(e)  & Mobility $\mu_h$(cm$^{2}$/Vs)              \\ \hline
H-Mo$_{9}$Se$_{18}$   & 1.84         & 77.86                      & T-Hf$_{9}$Se$_{18}$   & -7.60       & 5.26                                    \\ \hline
H-Mo$_{9}$Se$_{17}$S  & 1.80         & 81.70                      & T-Hf$_{9}$Se$_{17}$S  & -7.58       & 5.26                                     \\ \hline
H-Mo$_{9}$Se$_{17}$Te & 1.92         & 70.94                      & T-Hf$_{9}$Se$_{17}$Te & -7.73       & 5.26                                      \\ \hline
H-Mo$_{9}$Te$_{18}$   & 3.28         & 21.95                      & T-Zr$_{9}$S$_{18}$    & -7.26       & 5.26                                       \\ \hline
H-Mo$_{9}$Te$_{17}$S  & 3.17         & 23.66                      & T-Zr$_{9}$S$_{17}$Se  & -7.30       & 5.26                                         \\ \hline
H-Mo$_{9}$Te$_{17}$Se & 3.20         & 23.17                      & T-Zr$_{9}$S$_{17}$Te  & -7.40       & 5.26                                         \\ \hline
H-W$_{9}$Se$_{18}$    & 1.23         & 188.10                     & T-Zr$_{9}$Se$_{18}$   & -8.17       & 5.26                                        \\ \hline
H-W$_{9}$Se$_{17}$S   & 1.19         & 202.23                     & T-Zr$_{9}$Se$_{17}$S  & -8.13       & 5.26                                         \\ \hline
H-W$_{9}$Se$_{17}$Te  & 1.31         & 163.86                     & T-Zr$_{9}$Se$_{17}$Te & -8.28       & 5.26                                         \\ \hline
T-Sn$_{9}$S$_{18}$    & -4.89        & 5.26                       & T-Pt$_{9}$S$_{18}$    & -2.68       & 5.26                                         \\ \hline
T-Sn$_{9}$S$_{17}$Se  & -4.90        & 5.26                       & T-Pt$_{9}$S$_{17}$Se  & -2.66       & 5.26                                         \\ \hline
T-Sn$_{9}$S$_{17}$Te  & -4.80        & 5.26                       & T-Pt$_{9}$S$_{17}$Te  & -2.66       & 5.26                                         \\ \hline
T-Sn$_{9}$Se$_{18}$   & -4.98        & 5.26                       & T-Pt$_{9}$Se$_{18}$   & -2.36       & 5.27                                         \\ \hline
T-Sn$_{9}$Se$_{17}$S  & -4.98        & 5.26                       & T-Pt$_{9}$Se$_{17}$S  & -2.38       & 5.26                                        \\ \hline
T-Sn$_{9}$Se$_{17}$Te & -4.99        & 5.26                       & T-Pt$_{9}$Se$_{17}$Te & -1.96       & 5.54                                         \\ \hline
T-Hf$_{9}$S$_{18}$    & -6.88        & 5.26                       & T-Pt$_{9}$Te$_{18}$   & -1.56       & 20.54                                       \\ \hline
T-Hf$_{9}$S$_{17}$Se  & -6.91        & 5.26                       & T-Pt$_{9}$Te$_{17}$S  & -1.64       & 12.13                                       \\ \hline
T-Hf$_{9}$S$_{17}$Se  & -7.00        & 5.26                       & T-Pt$_{9}$Te$_{17}$Se & -1.61       & 14.53                                    \\ \hline

\end{tabular}}\label{table-defect hole mobility}
\end{table*}

In order to simulate the point defect, we expand the primitive cell of TMDs by 3$\times$3 to form a supercell with 27 atoms, then substituting one chalcogen atom (S, Se, Te) by another chalcogen atom to form a defective supercell, and then we calculate the BEC of these defective cells to obtain corresponding carrier mobility. According to the calculations, we found that the BEC of H-W$_9$Se$_{17}$S is 1.19 $|e|$, thus its electron mobility and hole mobility are 230.02 cm$^{2}$V$^{-1}$s$^{-1}$ and 202.23 cm$^{2}$V$^{-1}$s$^{-1}$, respectively, as shown in Table \ref{table-defect electron mobility} and Table \ref{table-defect hole mobility}. While for T-Zr$_9$Se$_{17}$Te, it has relative large BEC as 8.28 $|e|$, which result in related electron mobility and hole mobility of 25.32 cm$^{2}$V$^{-1}$s$^{-1}$ and 5.26 cm$^{2}$V$^{-1}$s$^{-1}$, respectively. We further found that the carrier mobility is related to chalcogen element (S, Se, Te) of TMDs. Generally, the order of the carrier mobility with the addition of chalcogen element is S$>$Se$>$Te. However, T-Pt$_9$X$_{17}$Y (X and Y are distinct element) is an exception, the order turns out inversely to be Te$>$Se$>$S. For example, the $|Z_B|$ of H-Mo$_{9}$Se$_{18}$ is calculated as 1.84 $|e|$. When one of its chalcogenides changes, its BEC changes simultaneously. The $|Z_B|$ of H-Mo$_{9}$Se$_{17}$S change to a smaller value of 1.80 $|e|$, and the $|Z_B|$ of H-Mo$_{9}$Se$_{17}$Te change to a bigger value of 1.92 $|e|$. The order of their $|Z_B|$ satisfy the rule as H-Mo$_{9}$Se$_{17}$S$<$ H-Mo$_{9}$Se$_{18}$$<$ H-Mo$_{9}$Se$_{17}$Te. The mobility of electron and hole of the H-Mo$_{9}$Se$_{17}$S is 112.36 cm$^{2}$V$^{-1}$s$^{-1}$ and 81.70 cm$^{2}$V$^{-1}$s$^{-1}$, and that of H-Mo$_{9}$Se$_{18}$ are 108.43 cm$^{2}$V$^{-1}$s$^{-1}$ and 77.86 cm$^{2}$V$^{-1}$s$^{-1}$, and that of H-Mo$_{9}$Se$_{17}$Te are 101.30 cm$^{2}$V$^{-1}$s$^{-1}$ and 70.94 cm$^{2}$V$^{-1}$s$^{-1}$, respectively, as shown in Table \ref{table-defect electron mobility} and Table \ref{table-defect hole mobility}.

Therefore, one want to find a new TMDs with high mobility, the BEC of the material should be small. The addition of S atoms can reduce the BEC, and obtain a relative small BEC, thus a large mobility of TMDs. On the contrary, for T-PtX$_{2}$ (X=S, Se and Te), the addition of Te results in a relative large mobility.

\section{Summary}
We have calculated the carrier mobility of pure and defective Janus structural materials using first-principles calculation and BEC. It is shown that the method of BEC is more suitable for calculating the carrier mobility of Janus TMDs than that of DPT. Our study reveals the reason of effective BEC, which is the carrier mobility of most Janus structural materials are mainly affected by longitudinal optical branch except H-MoSSe and H-WSSe. For the pure Janus structural materials, H-WSTe has the largest electron mobility of 129.72 cm$^{2}$V$^{-1}$s$^{-1}$ and hole mobility of 382.05 cm$^{2}$V$^{-1}$s$^{-1}$ due to the lowest BEC. While for Janus structural materials with defects, H-W$_9$Se$_{17}$S has the highest electron mobility of 230.02 cm$^{2}$V$^{-1}$s$^{-1}$ and hole mobility of 202.23 cm$^{2}$V$^{-1}$s$^{-1}$ due to the lowest BEC as well.

\section*{Conflicts of interest}
The authors declared that they have no conflicts of interest to this work.

\section*{acknowledgments}
The authors are grateful for financial support from National Natural Science Foundation of China (General program 51671086) and Hunan Provincial Education Department (Key project 19A324).

\end{document}